\documentclass[letter,twocolumn]{jpsj3}
\usepackage{graphicx,amsmath,amssymb,bm}
\usepackage{txfonts}

\allowdisplaybreaks 

\usepackage{color}
\definecolor{Green}{rgb}{0,0.7,0}

\newcommand{\bk}{ \bm{k}}

\newcommand{\bkD}{ \bm{k}_{\rm 0}}
\newcommand{\kDx}{ k_{\rm 0x}}
\newcommand{\kDy}{ k_{\rm 0y}}
\newcommand{\kDz}{ k_{\rm 0z}}

\begin{document}
\title{
Effective Hamiltonian 
 of Topological Nodal Line Semimetal 
 in Single-Component Molecular Conductor 
[Pd(dddt)$_2$] from First-Principles
}
\author{
 Takao Tsumuraya$^{1}$ 
\thanks{E-mail: tsumu@kumamoto-u.ac.jp}, 
Reizo Kato$^{2}$ 
\thanks{E-mail: reizo@riken.jp}, 
and 
Yoshikazu Suzumura$^{3}$
\thanks{E-mail: suzumura@s.phys.nagoya-u.ac.jp}
}
\inst{
$^1$
Priority Organization for Innovation and Excellence, 
Kumamoto University, 
 Kumamoto~860-8555, Japan \\
$^2$
 Condensed Molecular Materials Laboratory, RIKEN,  Wako, Saitama 351-0198, Japan \\
$^3$
Department of Physics, Nagoya University,  Nagoya 464-8602, Japan }
\recdate{August  21, 2018, Accepted September 18, 2018, Published online October 10, 2018}
\abst{
Using first-principles density-functional theory calculations,
we obtain the non-coplanar nodal loop for a single-component molecular conductor [Pd(dddt)$_2$] consisting of HOMO and LUMO with different parity. 
Focusing on  two typical Dirac points,
we present a model of an effective 2 $\times$ 2 matrix Hamiltonian in terms of 
  two kinds of velocities associated with the nodal line.
The base of the model is taken as HOMO and LUMO on each Dirac point, 
 where two band energies degenerate and the off diagonal matrix element vanishes.
 The present model, which reasonably describes  the Dirac cone 
 in accordance with the first-principles calculation,  provides a new method of analyzing electronic states of a topological nodal line semimetal.
}
\maketitle
Single-component molecular conductor, [Pd(dddt)$_2$], which is insulating at ambient pressure but becomes conducting and shows almost temperature independent resistivity under a pressure of 12.6 GPa,~\cite{Kato_JACS} has been studied using first-principles density-functional theory~(DFT) calculations,~\cite{Kohn_Sham}  and linearly dispersed bands with Dirac cones were found at 8~GPa.~\cite{Kato_JACS, Tsumuraya2018} 
An extended H\"uckel calculation and a tight-binding analysis of the DFT optimized structure at 8 GPa have shown nodal line semimetal.~\cite{Kato2017_JPSJ}. 
Relevant physical quantities have been examined to understand 
 the characteristics of such
new massless Dirac electron systems.~\cite{Suzumura_Pddddt2_JJAP2017, Suzumura_Conduc_2017, Suzumura2018_JPSJ_T, Suzumura_Yamakage_JPSJ} 
In fact, the  multi-orbital nature due to an interplay between HOMO (highest occupied molecular orbital) and LUMO (lowest unoccupied molecular orbital) on different molecular sites is essential  
 for the emergence of the Dirac electron.~\cite{Kato_JACS} 
Such a state, where the contact points of the Dirac cones form a closed line (loop) called nodal line semimetal,~\cite{Kim_Rappe_NLS_2015, Weng_Line_node, KieranPRL, Fang_NLS, Liu_Balents2017}
 is quite different from that of massless Dirac electrons in a two-dimensional molecular conductor.~\cite{Katayama2006_JPSJ75}

In 1937, Herring~\cite{Herring_PR.52.365} 
 claimed accidental degeneracies in the energy bands
  followed by a closed/open circuit (nodal line) for a system with inversion symmetry.
 The product of parity eigenvalues at the eight time reversal invariant momentum (TRIM)  determines  even or odd number of circuits (nodal loop). 
This formula is quite similar to Fu-Kane's criteria of $\mathbb{Z}_2$ topological insulator,~\cite{Fu_Kane_PRB.76.045302} despite assuming the absence of spin-orbit coupling (SOC). 
However, since it  should work well for weak SOC materials with 
 light elements such as molecular conductors, 
 the shape of the nodal line and the number of nodal loops would be useful to 
evaluate the band topology and design new molecular topological materials.  
The  nodal line semimetal  is a recent topic  
related to the new concept associated with the topological property of Dirac electrons.~\cite{Murakami2007,Burkov2011,Hirayama2017,Hirayama2018} 
Among them, single nodal loop has been found 
in pnictides CaAg$X$ ($X$ = P and As)\cite{Yamakage_CaAgX_Dirac}, alkaline-earth compounds of $A$$X_2$ ($A$ = Ca, Sr, Ba; $X$ = Si, Ge, Sn),~\cite{Liu2016PRB} CaAs$_3$\cite{Diracloop_CaAs3_2017}, and Ag$_2$S~\cite{Ag2S_PRB.96.115106}. In fact, all these systems and [Pd(dddt)$_2$] turn out to be a strong topological insulator\cite{Fu_Kane_Mele2007} when SOC is on.~\cite{Tsumuraya2018, Kato2017_JPSJ, Liu2018} 

With the regard to the Dirac electrons in [Pd(dddt)$_2$],
 it is useful to describe the nodal line semimetal in terms of the effective Hamiltonian with a two-band model. 
For many nodal line semimetals, first-principles calculations have characterized the shape of the 
 nodal line, and an effective Hamiltonian is analyzed only close to the 
$\Gamma$ point due to perturbation.~\cite{Liu2016PRB, Liu2018}
However, such a  Hamiltonian is insufficient to describe   
  the large nodal line away from TRIM due to the non-coplanar~\cite{Kato2017_JPSJ} 
and snake-like loop structures.~\cite{Liu2016PRB} 
Further, the model is required to show    
 a reasonable energy band for the Dirac cone along the nodal line
 in order to examine the physical property, 
 since the chemical potential (i.e., the Fermi energy) crosses the nodal line. 
Indeed,  the model, which provides 
 two kinds of velocity fields associated with the Dirac cone, is useful as shown by  the calculation of the topological property of the  Berry phase.
 \cite{Suzumura_Yamakage_JPSJ} 
In single-band molecular conductors, first-principles band structures near the Fermi level 
are well reproduced by a small number of hopping parameters,~\cite{Kino2006_JPSJ, Koretsune_Hotta_PRB2014} 
but the corresponding parameters for a multi-band system of [Pd(dddt)$_2$] are huge and complicated.
  Therefore, a new method for the direct derivation of the model 
   from the  DFT calculation is needed to describe  
 the Dirac cone reasonably for all the Dirac points along the loop. 

In this letter, on the basis of first-principles DFT calculations, 
  we calculate a number of  Dirac points that form 
   a  single non-coplanar nodal loop  centered at the $\Gamma$-point 
    in  the first Brillouin zone (BZ). 
 The characteristics of the two crossing bands near the Dirac points 
 and the velocities of the Dirac cone  
 are examined by focusing on two typical Dirac points on the loop. 
 Then, we demonstrate a new method to obtain the effective Hamiltonian for 
the nodal line semimetal 
 where  the base  is not on the TRIM, but on the respective 
   Dirac point.~\cite{Luttinger1955,Kobayashi2007_JPSJ}
 A model with analytical matrix elements
 is derived using the knowledge of  
 the general relation between the nodal line 
 and the velocity fields of the Dirac cone.\cite{Suzumura_Yamakage_JPSJ}  

The crystal structure of [Pd(dddt)$_2$], which is determined by x-ray diffraction at ambient pressure,  belongs to a monoclinic system 
 with the space group of $P2_1$/$a$.\cite{Kato_JACS} 
In our previous study, using first-principles calculations, we discovered anisotropic Dirac cones of [Pd(dddt)$_2$] at 8GPa by performing structural optimization under symmetry operations.~\cite{Kato_JACS}. 
However, the three-dimensional characteristics of the electronic state and the effective model have not yet been reported  from first-principles. 
The present first-principles DFT calculations are performed based on an all-electron full-potential linearized plane wave (FLAPW) method as implemented in QMD-FLAPW12.~\cite{Wimmer1981, L_KA, Weinert} 
To  determine accurately the anisotropy of the Dirac cone, Fermi velocities, and contact points of Dirac cones along the nodal line, we calculate the eigenvalues around the Dirac cone  by using the highly precise FLAPW method. We used a high-dense $\bk$-mesh for plotting the 3D band structures.  
The exchange-correlation functional that is used in the present calculation is the generalized gradient approximation proposed by Perdew, Burke, and Ernzerhof.~\cite{GGA_PBE} The $\bk$-point sampling grid of 
4 $\times$ 8 $\times$ 4 was used. 
The BZ integration was performed with an improved tetrahedron method.~\cite{Imp_Tetrahedron1994}

\begin{figure}
  \centering
\includegraphics[width=1.0\linewidth]{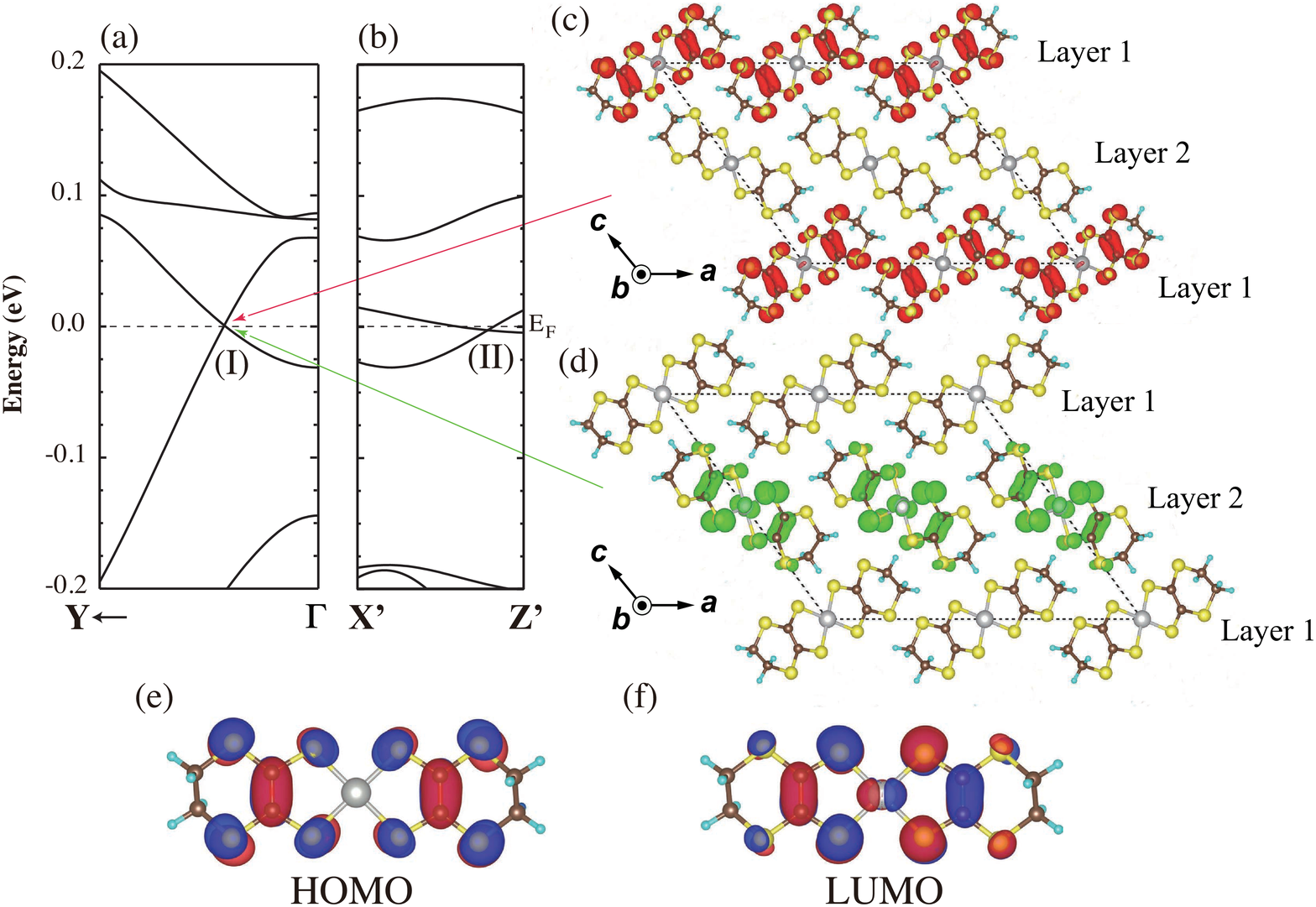} 
    \caption{(Color online)
Band structure of [Pd(dddt)$_2$] at 8GPa (without SOC) along the $\Gamma$-Y direction (a) and  along the X'(-0.1967, 0, 0)-Z'(-0.1967, 0, 0.5) direction (b) 
 on which Dirac points (I) and (II) are found, respectively. 
Wavefunction at a fixed $\bk$: charge densities $|\Psi_{i,\bk}|^2$ close to  Dirac point (I) for the bottom of the conduction bands (c), 
and those at the top for the valence bands  (d). 
The notations \textsl{\textbf{a}}, \textsl{\textbf{b}}, and \textsl{\textbf{c}} denote  the respective 
 lattice vectors of the unit cell.  
Contour plot of the orbital (charge) density of the HOMO (e) and  LUMO (f) 
 of isolated [Pd(dddt)$_2$] monomer. 
 }
\label{fig:structure}
\end{figure}

Figures \ref{fig:structure}(a) and \ref{fig:structure}(b) show  
the band structure along the $\Gamma$-Y and 
  along X'(-0.1967, 0, 0)-Z'(-0.1967, 0, 0.5) directions, respectively 
  where the band crossing suggests 
two typical Dirac points.
We refer to  the former Dirac point as  Dirac point (I) 
 and the later as  Dirac point (II). 
In general,
   the crossings of two linear bands with different characteristics
occurs forming a Dirac cone. 
By comparing the characteristics  of the HOMO and LUMO for the isolated [Pd(dddt)$_2$] monomer 
in Figs.~\ref{fig:structure}(e)  and \ref{fig:structure}(f),
 we discuss the wavefunction near  Dirac point (I). 
As plotted in Figs.~\ref{fig:structure}(a)--\ref{fig:structure}(d), 
 the convex upwards band 
   predominately consists of HOMO like wavefunctions in  Layer 1, and 
the convex downwards band is made up of LUMO like wavefunctions in  Layer 2.

\begin{figure}
  \centering
\includegraphics[width=0.9\linewidth]{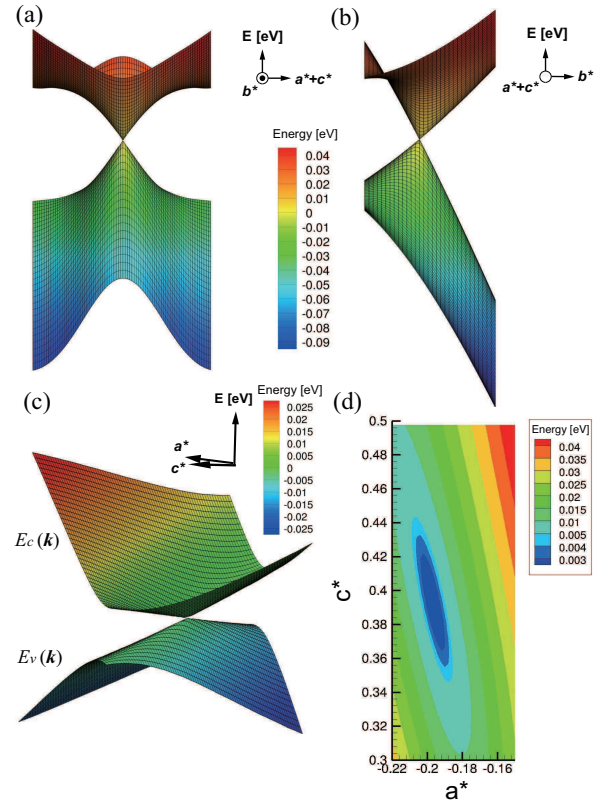}
    \caption{(Color online) Band structure 
  close to  Dirac point (I) on the $k_{x+z}$-$k_y$ plane,  
 where (a) and (b) are 
the  side views with the horizontal axis given by 
 \textsl{\textbf{b*}} with $0.06~<~k_y~<~0.122$   and  
  by \textsl{\textbf{a*}}+\textsl{\textbf{c*}} with $-0.2 < k_{x+z} < 0.2$, 
respectively. 
 (c) Bird's eye view of the Dirac cone 
  with apex at Dirac point (II) on the $k_x$-$k_z$ plane,  
  where $-0.22 < k_x < -0.15$ and $0.3 < k_z <  0.5$.  
 (d)  2D contour plots corresponding to (c) 
 for the difference in energy between the top of the valence bands, $E_v$($\bk$), and the bottom of conduction bands, $E_c$($\bk$),
 on the \textsl{\textbf{a*}}-\textsl{\textbf{c*}} plane,     
 where $k_x$ and $k_z$ represent 
 \textsl{\textbf{a*}} and \textsl{\textbf{c*}}, respectively,  which  are taken  as the orthogonal axes, for simplicity. 
 }
\label{fig:cone}
\end{figure}

The shape of the Dirac cone 
 depends on the position of the Dirac point along the nodal line.
In the following, we use a wavevector scaled by $2\pi$,   
   lattice constants of the reciprocal lattice taken as unity, 
 and energy in the unit of eV. 
Figures \ref{fig:cone}(a) and \ref{fig:cone}(b) show the Dirac cone with  
   Dirac point (I), $\bkD$ = (0, 0.086, 0), where
  the axis of the cone is parallel to 
a vector (--2$^{-1/2}$, 0, 2$^{-1/2}$). 
The principal axes of the plane, which are perpendicular to 
 the axis of the Dirac cone,
are given by \textsl{\textbf{a*}}+\textsl{\textbf{c*}} ($k_{x+z}$) and
   \textsl{\textbf{b*}} ($k_y$).  
 Dirac cone (I) shown in Fig.~\ref{fig:cone}(a) is symmetric with respect to $k_{x+z}$ while 
 that shown in Fig.~\ref{fig:cone}(b) is tilted along the positive $k_y$.
 Figures \ref{fig:cone}(c) and (d) show the Dirac cone  of  Dirac point (II), 
 $\bkD$ = (-0.1967, 0.000, 0.3924), 
 where the axis of the cone is parallel to \textsl{\textbf{b*}} ($k_y$).
The former shows a bird's eye view with tilting along  both 
  $k_x$ and $k_z$ axes.
The latter denotes the gap function around Dirac point (II),  
which is the difference in energy between the valence and  conduction bands, $E_c(\bk)-E_v(\bk)$. 
The principal axes of the Dirac cone (II) is rotated counterclockwise
 with an angle $\theta \sim \tan^{-1}(0.2)$  and 
  the  tilting does not appear.
The difference in behavior of
the  Dirac cone between Dirac points (I) and (II) 
is discussed in terms of the effective Hamiltonian.

\begin{figure}
  \centering
\includegraphics[width=8cm]{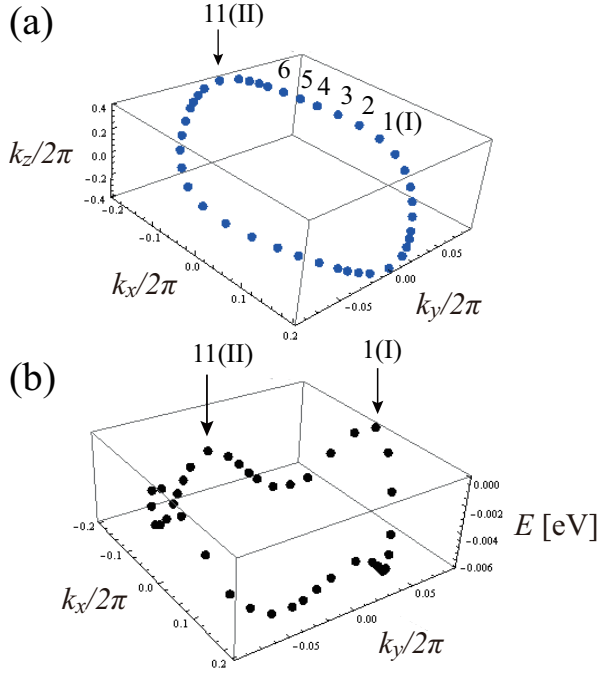}
    \caption{(Color online)
(a) Nodal line of accidental degeneracies in [Pd(dddt)$_2$] at 8GPa. The loop  is formed by the trajectory of the contact points of Dirac cones in the three-dimensional momentum space, $(k_x, k_y, k_z)$ in the absence of spin-orbit coupling.
The locations of the respective  points on the line,  which 
 are scaled by $2\pi$ 
$\bkD [= (\kDx, \kDy, \kDz)]$ are given by 
(0, 0.086, 0) (1),
(-0.042, 0.083, 0.05) (2),
(-0.078, 0.0755, 0.1) (3),
(-0.109, 0.066, 0.15) (4),
(-0.131, 0.057, 0.2) (5),
(-0.151, 0.047, 0.25) (6),
(-0.168, 0.037, 0.3) (7),
(-0.176, 0.0315, 0.325) (8),
(-0.184, 0.0245, 0.35) (9),
(-0.191, 0.0165, 0.375) (10), and 
(-0.1967, 0.000, 0.3924) (11)
where we mainly examine the symmetric points 
 (1) and (11) defined by (I) and (II) respectively.
(b) The corresponding energy 
 dispersion of the nodal line, $E(\bk)$ [eV]. The origin of the energy is set to be the energy at Dirac point (I). 
 The energy is the same for Dirac points shown  by 
 $\bkD$ = $(\kDx, \pm \kDy, \kDz)$  and 
 $(-\kDx, \pm \kDy, -\kDz)$. 
 }
\label{fig3}
\end{figure}


Figure \ref{fig3}(a)
 shows the nodal line (trajectory of Dirac points) of [Pd(dddt)$_2$] 
 at 8~GPa, which consists of 11 independent Dirac points $\bkD$ calculated 
 using the first-principles calculations. 
 Note that the present DFT calculations show a  
 non-coplanar loop  within the first BZ, while 
 the loop of the previous tight-binding model~\cite{Kato2017_JPSJ} 
  extends over the first BZ.
  For the estimation of the effective Hamiltonian, we focus on Dirac points (I) and (II) represented by the numbers 1 and 11, respectively. 
  The trajectory of the Dirac points suggests an accidental degeneracy, 
 which occurs without the symmetry of the crystal. 
 Note that these Dirac points are symmetric with respect to the $\Gamma$-point and the plane 
 of $k_y=0$~($k_x$--$k_z$ is a mirror plane), respectively. 
  
Using  Fig.~\ref{fig3}(a), we examine the effective Hamiltonian 
 of a two-band model, 
 where the base is given by the respective Dirac point $\bkD$. 
Note that this method describes the Dirac cone correctly, 
  as shown in Refs.~\citen{Luttinger1955} and ~\citen{Kobayashi2007_JPSJ}.
Two energies consisting of 
different kinds of parities degenerate on the Dirac point, 
 where arbitrary linear combination of these two wavefunctions 
 becomes possible.
 The eigenfunctions with $\bk$, which is  just away from the Dirac point 
 is determined uniquely, 
and their linear combination rotates around the Dirac point.\cite{Katayama2009_EPL} 
 In the present case, we can choose two eigenfunctions  of 
  HOMO and  LUMO  with different symmetry at the Dirac point.

\begin{figure}
  \centering
\includegraphics[width=1.0\linewidth]{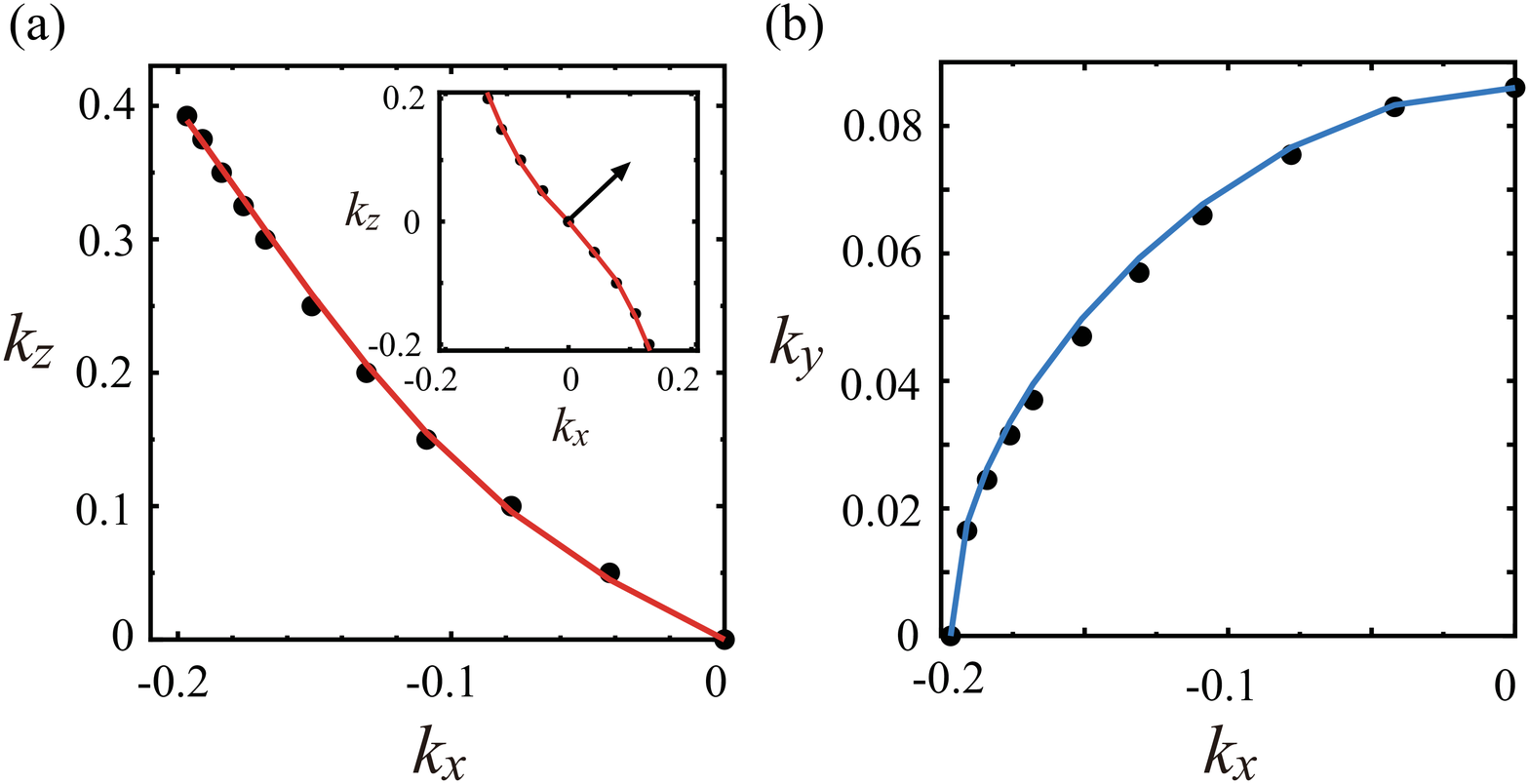}
    \caption{(Color online) Dirac points (symbols) projected on the $k_x$ - $k_z$ plane (a) and 
   $k_x$ - $k_y$ plane (b), which are used to obtain 
  the fitting lines 
for $f_2(\bk)$ and $f_3(\bk)$ 
  (Eqs.~(\ref{eq:eq7a}) and (\ref{eq:eq7b})), respectively.
 The inset shows $f_2(\bk)$ 
  around the $\Gamma$-point.  
 The arrow, which  depicts  $\bm{v}_2$ at 
   Dirac point (I) with arbitrary scale,   
   corresponds 
 to  the horizontal axis \textsl{\textbf{a*}}+\textsl{\textbf{c*}} 
in Fig.~\ref{fig:cone}(a).
}
\label{fig4}
\end{figure}

Now we derive the effective Hamiltonian of the two band model  explicitly.
We start with the full 8 $\times$ 8 Hamiltonian that is  
  represented at the $\Gamma$-point. Such a Hamiltonina 
  is given by 
\begin{eqnarray}
  H(\bk) 
 &=& 
\begin{pmatrix}
H^{\rm HH}(\bk)   &  H^{\rm HL}(\bk)  \\
H^{\rm LH}(\bk) & H^{\rm LL}(\bk) 
\end{pmatrix} \ , 
\label{eq:Heff}
\end{eqnarray}
  where the  4 $\times$ 4 matrix $H^{\rm HH}(\bk)$ ( $H^{\rm LL}(\bk)$) consists of only HOMO (LUMO) and $H^{\rm HL}(\bk)$ ($H^{\rm HL}(\bk)$) denotes 
 the interaction between HOMO and LUMO.
The former is the even function of $\bk$ due to the time reversal symmetry,  
 while the latter is the odd function due to interaction with 
 different parity.
Among  the energy bands of the four HOMOs and four LUMOs,  
we choose 
 two bands given by the top HOMO bands and the bottom  LUMO bands, 
 which are assumed to be isolated from the other bands. 
The  Scr\"odinger equation for these two bands 
$E_{\rm H}(\bk$) and $E_{\rm L}(\bk)$ is given by 
\begin{eqnarray}
 H^0(\bk) |H (L)(\bk)> =  E_{\rm H (L)}(\bk) |H (L)(\bk)> \; , 
\end{eqnarray}
 where $H^0({\bk})$ denotes Eq.~(\ref{eq:Heff}) 
 without  $H^{\rm HL}(\bk)$  and  $H^{\rm LH}(\bk)$.  
By using $|H (\bk)>$ and $|L (\bk)>$, 
 the reduced Hamiltonian of the 2 $\times$ 2 matrix is written as 
\begin{eqnarray}
 H^{\rm  red} & =& \sum_{\alpha, \beta = {\rm H}, {\rm L}}
   |\alpha(\bk)>  <\alpha(\bk)| 
  H(\bk)   |\beta(\bk)>  <\beta(\bk)| \nonumber  \\
& & = \sum_{\alpha, \beta = {\rm H}, {\rm L}} |\alpha(\bk)>  H_{\rm eff}  <\beta(\bk)| \; .
\label{eq:eq3}
\end{eqnarray}
 By defining 
 \begin{eqnarray}
 f_0 &=&  (E_{\rm H}(\bk) + E_{\rm L}(\bk)|/2 \; , 
                                       \nonumber \\ 
 f_3 &=&  |E_{\rm H}(\bk) - E_{\rm L}(\bk)|/2 \; , 
                                       \nonumber \\ 
 f_2 &= & i <{\rm H}(\bk)| H |{\rm L}(\bk)>\; , 
\label{eq:eq4}
\end{eqnarray}
$H_{\rm eff}(\bk)$ in Eq.~(\ref{eq:eq3}) is expressed as
\begin{eqnarray}
 {H_{\rm eff}(\bm{k})}  
 &=& 
\begin{pmatrix}
f_0(\bk) + f_3(\bk) & - i f_2(\bk)  \\
i f_2 (\bk) & f_0(\bk)- f_3(\bk) 
\end{pmatrix} \ , 
\label{eq:eq5}
\end{eqnarray}
where $\bk = (k_x, k_y, k_z)$ and  $f_0(0)$ is taken as zero.  
  The energy is given by 
 $E_{\pm} =f_0 \pm \sqrt{f_2^2 + f_3^2}$, 
  where  $E_+ (E_-) = E_c (E_v)$. 
 The Dirac point $\bkD$ on  the nodal line is obtained from 
\begin{subequations}
\begin{eqnarray}
f_2(\bkD) &=& 0 \; , 
\label{eq:eq6a}
\\
f_3(\bkD) &=& 0 \; .
\label{eq:eq6b}
\end{eqnarray}
\end{subequations} 
Note that $f_0(\bk)$ and $f_3(\bk)$ are
 the even functions of $\bk$ due to time reversal symmetry. 
The quantity  $f_2(\bk)$,  which is also real due 
 to the presence of  inversion symmetry   around the Pd atom,~\cite{Kato2017_JPSJ}
is the odd function of $\bk$ due to 
 HOMO and LUMO having different parities. 
The function $f_2(\bk)$  is estimated  by projecting the nodal line on the  $k_x$ - $k_z$ plane in   Fig.~\ref{fig4}(a), while 
 $f_3(\bk)$  is estimated by projecting the nodal line on the $k_x$ - $k_y$ plane in  Fig.~\ref{fig4}(b). 
This method reproduces the Dirac points  well as shown in Fig.~\ref{fig3}(a).
 We empirically examined the form under the constraint that 
the Dirac points  (0,~0.086,~0) (I) and 
 (-0.1967, 0.000, 0.3924) (II) satisfy Eqs.~(\ref{eq:eq6a})~and (\ref{eq:eq6b}).
These functions  are obtained as 
\begin{subequations}
\begin{eqnarray}
f_2(\bk) & \simeq & C_2(k_z +  k_x + 40 k_x^3 - 380 k_x^5) \; ,
\label{eq:eq7a}
     \\
f_3(\bk) & \simeq & C_3((k_x/0.1967)^2 + (k_y/0.086)^2 
   \nonumber \\
   & &  +(k_xk_y/0.027)^2 - 1 )  \; , 
\label{eq:eq7b}
 \\
f_0(\bk)& \simeq &  b_xk_x^2 + b_yk_y^2 + b_zk_z^2 + C_0\; .
\label{eq:eq7c}
\end{eqnarray}
\end{subequations}
 where 
 the velocity of the cone at the Dirac point
 $\bkD$ is obtained
 as $\bm{v}_2  =\nabla_{\bkD}f_2$, $\bm{v}_3  =\nabla_{\bkD}f_3$ and   
   $\bm{v}_0  =\nabla_{\bkD}f_0$ corresponding to  the tilting. 
Using  $\bkD = (\kDx, \kDy, \kDz)$, $\bm{v}_j$ 
 is calculated as 
\begin{subequations}
\begin{eqnarray}
 \bm{v}_2 & = & \nabla_{\bkD}f_2  \simeq  
         C_2(1 + 120 \kDx^2 - 2280\kDx^4, 0, 1), 
                        \label{eq:eq8a}             \\
 \bm{v}_3 & = & \nabla_{\bkD}f_3   \simeq  
       C_3(2\kDx/0.196^2+2\kDx\kDy^2/0.027^2, 
                                     \nonumber \\
      & &2\kDy/0.086^2 + 2\kDx^2\kDy/0.027^2, 0)
                               \; ,  \label{eq:eq8b} \\
 \bm{v}_0 & = & \nabla_{\bkD}f_0 \simeq (2b_x\kDx,2b_y\kDy, 2b_z\kDz) \; .
   \label{eq:eq8c} 
 \end{eqnarray}
\end{subequations}
Note that the tangent of the nodal line is parallel to  
 $\bm{v}_2 \times \bm{v}_3$.\cite{Suzumura_Yamakage_JPSJ}
In the previous work,\cite{Kobayashi2007_JPSJ}
 $\bm{v}_j$ was estimated directly from Eq.~(4) 
    using the wavefunction close to  $\bk=\bkD$. 
 From $f_j(\bk) - f_j(\bkD) \simeq \bm{v}_j \cdot \delta \bm{k}$,  ($j$ = 2, 3, 0)  with  $\delta \bk = \bk - \bkD$, 
 the energy  of the  Dirac cone is given by 
 $E_{\pm}(\bk) -  E(\bkD) \simeq  \bm{v}_0 \cdot \delta \bk$ 
 $\pm 
 \sqrt{ (\bm{v}_2 \cdot \delta \bk)^2 + (\bm{v}_3 \cdot \delta \bk)^2}$.
 The coefficients of $k_x^3$, $k_x^5$ and $(k_xk_y)^2$ 
 in Eqs.~(\ref{eq:eq7a}) and (\ref{eq:eq7b})
 are determined    by fitting the line to the other $\bkD$. 
 As shown in Figs.~\ref{fig4}(a) and \ref{fig4}(b),
 good coincidence is obtained for both $f_2(\bk)$ and  $f_3(\bk)$ where 
 $f_2(\bk)$ reflects the symmetry of $k_y$=0. 
 The arrow in the inset of Fig.~\ref{fig4}(a) represents  $\bm{v}_2$ at Dirac point
 (I) which is perpendicular to $f_2(\bk)$ (solid line).
 Coefficients  $C_2$, $C_3$, $b_x$, $b_y$, and $b_z$ 
 are calculated  as follows.

First, we examine the Dirac cone at  Dirac point (I) 
 where 
$\bm{v}_2 = C_2(1,0,1)$, 
$\bm{v}_3 = C_3(0,1,0)$, and  
$\bm{v}_0 = 2 b_y \kDy (0,1,0)$. 
Noting  $\bm{v}_2 \cdot \bm{v}_3$ =~0,
  the principal axes are given by
   $k_x$+$k_z$ and  $k_y$.
 By comparing   the velocities  with 
those of  Figs.~\ref{fig:cone}(a) 
 and \ref{fig:cone}(b), which give  
\begin{subequations}
\begin{eqnarray}
v_2 &=&  \bm{v}_2 \cdot (2^{-1/2}, 0, 2^{-1/2}) = 0.21  \; , 
  \label{eq:eq9a}             \\
 v_y &=& \bm{v}_3 \cdot (0, 1,0) \pm \bm{v}_0 \cdot (0,1,0)  = 1.25 \pm 0.45
 \; ,  \label{eq:eq9b} 
    \nonumber \\
 \end{eqnarray}
\end{subequations}
 we obtain $C_2 \simeq 0.148$, $b_y \simeq - 2.6$, and $C_3  \simeq 0.053$.  
Coefficients $C_2$ and $C_3$ can also be  obtained  
 from  the gap $E_c(\bk)-E_v(\bk)$ at the $\Gamma$-point and Z-point,
 which are estimated as 0.0994 and 0.1117, respectively.
The resultant quantities    
$C_2$ = 0.137 and $C_3$ = 0.050 are compatible with those of 
Eqs.~(\ref{eq:eq9a}) and (\ref{eq:eq9b}), 
 which suggests that the effective Hamiltonian of Eq.~(\ref{eq:eq5}) 
 is valid not only for $\bk$ close to the nodal line but also for the 
extended regions including the $\Gamma$-  and Z-points.

  Next, we examine the Dirac cone at  Dirac point (II) 
to verify the validity of Eqs.~(\ref{eq:eq7a}) and  (\ref{eq:eq7a}) 
 with $C_2 =0.148$ and  $C_3 = 0.053$.
 The  Dirac cone is shown  on the plane of $k_x$ - $k_z$, 
 but the principal axes do not coincide with that of $\bm{v}_2$ and $\bm{v}_3$ 
  since $\bm{v}_2 \cdot \bm{v}_3 \not= 0$
  from Eqs.~(\ref{eq:eq8a}) and (\ref{eq:eq8a}). 
 We estimate the gap function  given by 
$ 2 \Delta (\bk) = E_c(\bk)-E_v(\bk)$
$ = 2\sqrt{ (\bm{v}_2 \cdot \delta \bk)^2 + (\bm{v}_3 \cdot \delta \bk)^2}$
 with  $\delta \bk =(x,0,z)$, which is calculated as  
 $(E_c(\bk)-E_v(\bk))^2/4 = 0.45 x^2 + 0.117 x z + 0.02 z^2$
 $(= Ax^2 + 2C xz + Bz^2)$. 
 This shows the rotation of the principal axes  
 from $(x,z)$ to $(X,Z)$, where 
$x = X \cos \theta - Z \sin \theta$, and  
$z= X \sin \theta  + Z \cos \theta$. In terms of 
 $(X,Z)$, the gap function is expressed as  
\begin{eqnarray}
 & & \Delta(\bk)^2 = \Delta(\delta \bk +\bkD) = A x^2  + 2C xz + B z^2  =   
  \nonumber \\ 
  & &  \frac{1}{2}\left( A+B + \sqrt{(A-B)^2 + 4C^2}\right) X^2 
                \nonumber \\
   & &  + \frac{1}{2}\left( A+B - \sqrt{(A-B)^2 + 4C^2}\right) Z^2
 \; ,
    \nonumber \\ 
& & \tan (2 \theta) = \frac{2C}{A-B} \; .
\label{eq:eq10}
\end{eqnarray}
The cross section of $\Delta(\bk) = E_0 $  
 shows an ellipse   with the radius of the  minor [major] axis given by 
  $a=2^{1/2}E_0/\sqrt{A+B+((A-B)^2+4C^2)^{1/2}}$
 [$b$ = $2^{1/2}E_0/\sqrt{A+B-((A-B)^2+4C^2)^{1/2}}$],
 which is 
rotated by an angle $\theta = 2^{-1}\tan^{-1}[2C/(A-B))]$. 
The radius and angle for Dirac point (II) 
 are obtained as  $a = 1.5 E_0$, $b=8.5 E_0$, $b/a \simeq 5.6$ and $\tan \theta = 0.13$, which 
 correspond well to  the numerical result in  Fig.~2(d), i.e.,  
    $b/a \simeq 7$ and $\tan \theta \simeq 0.2$. 
 Therefore,  our Hamiltonian may  be applied to all the Dirac points 
 between points (I) and (II).

Further, from  Eq.~(\ref{eq:eq10}),  we  estimate    
   the area of the ellipse $S$ with the gap $2 E_0$ 
  for an arbitrary Dirac point, where  
 the axis of the cone is parallel to the tangent of the nodal line 
( i.e.,  $\propto \bm{v}_2 \times \bm{v}_3$) 
 and the wavefunction  of the Dirac cone 
 is determined by  $\bm{v}_2$   and  $\bm{v}_3$.  
By taking the axis of the cone as $y$ and the plane of the cone 
 as the  $x$ - $z$ plane,
  we obtain the area of the ellipse $S$ with the gap $2 E_0$  
 as 
 $S(\bkD) = \pi ab$  $= \pi E_0^2/ \sqrt{AB - C^2} 
 = \pi E_0^2 / |\bm{v}_2(\bkD) \times \bm{v}_3(\bkD)|$,
  where 
$\bm{v}_2 = (v_{2x}, 0, v_{2z})$, 
$\bm{v}_3 = (v_{3x}, 0,  v_{3z})$,  
$A= v_{2x}^2+v_{3x}^2$, 
$B= v_{2z}^2+v_{3z}^2$, and 
$C=v_{2x}v_{2z} v_{3x}v_{3z}$. 
We note that 
$\bm{v}_2 \cdot \bm{v}_3 \not= 0$ 
is generally  expected  
 except for  Dirac point (I), which is located 
on the symmetric line of  $\Gamma$ - Y.

Finally, we calculate 
 $b_x$ and $b_z$ in Eq.~(\ref{eq:eq7c})  
 from  Dirac cone (II), 
which gives $v_x = 0.36 \pm 0.12$ and $v_z = 0.09 \pm 0.06$. 
In a way similar to Eqs.~(\ref{eq:eq9a}) and (\ref{eq:eq9b}), 
 we obtain $b_x \simeq -0.30$ and $b_z \simeq 0.077 $   
 suggesting that the tilting occurs 
 towards   the negative (negative) direction  for the $k_x$ ($k_z$) 
axis. 
  This is  qualitatively consistent with Fig.~2(c). 
  As shown in Fig.~\ref{fig3}(b), 
     our DFT calculation of the energy difference  
  measured from that of  Dirac point (I) exhibits 
   nonmonotonical behavior, e.g.,  
 -0.0031, -0.0061, -0.0059, -0.0047,  and -0.0032~eV 
 for the nodal points  No. 3, 5, 7, 9, and 11 in Fig.~\ref{fig3}(a), respectively.
 However, it is quite complicated to reproduce the dispersion
 quantitatively from Eq.~(\ref{eq:eq7c}) with a constant $C_0$, 
and  such a problem becomes an issue to be solved in the next step.

Here, it should be noted that  the SOC  was ignored in the present calculation.
Since the  crystal structure is centrosymmetric, 
  the band structures are in Kramers degeneracy. 
Then, the loop degeneracies are destroyed by the SOC,  
 leading to  an energy gap with $\simeq$  3 meV  at Dirac point (I),~\cite{Tsumuraya2018} which will be examined further.

We comment on the experiment in which the present model is useful 
 for performing the analysis.  
Since all the directions for the nodal line can be treated 
   on the same footing, 
the effect of magnetic field is of interest  to examine the characteristic 
 behavior of the Landau level. The Hall coefficient  is 
 also expected to exhibit anisotropic  behavior.

In summary, we performed first-principles calculations for [Pd(dddt)$_2$] at 8GPa, and found a non-coplanar nodal loop within the first BZ. 
 The present effective Hamiltonian obtained 
 from the Dirac points in Fig.~\ref{fig3}(a) 
 describes  well the electron  close to the nodal line.
 The explicit form of the velocity is obtained  by  
 Eqs.~(\ref{eq:eq8a})-(\ref{eq:eq8c}), with 
 $C_2 \simeq 0.148, 
C_3 \simeq 0.05, 
b_x \simeq -0.31, 
b_y \simeq - 2.6$, and  
$b_z \simeq 0.077$,
 which are useful for  future calculations of 
transport.

\acknowledgements
The authors thank T. Miyazaki, H. Kino, A. Yamakage, F. Ishii, H. Sawahata, T. Shishidou, and T. Kariyado for useful discussions. 
The computations were  mainly carried out using the computer facilities of ITO 
  at the Research Institute for 
Information Technology, Kyushu University and HOKUSAI-GreatWave at RIKEN. 
This work was supported by JSPS KAKENHI Grant, JP16H06346 and 16K17756.  

%

\begin{thebibliography}{10}

\bibitem{Kato_JACS}
R.~Kato, H.~Cui, T.~Tsumuraya, T.~Miyazaki, and Y.~Suzumura, J. Am. Chem. Soc.
  {\bfseries 139}, 1770 (2017).

\bibitem{Kohn_Sham}
W.~Kohn and L.~J. Sham, Phys. Rev. {\bfseries 140}, A1133 (1965).

\bibitem{Tsumuraya2018}
T.~Tsumuraya, H.~Sawahata, F.~Ishii, H.~Kino, R.~Kato, and T.~Miyazaki, Am.
  Phys. Soc. Bull, R14.00012 (2018).

\bibitem{Kato2017_JPSJ}
R.~Kato and Y.~Suzumura, J. Phys. Soc. Jpn. {\bfseries 86}, 064705 (2017).

\bibitem{Suzumura_Pddddt2_JJAP2017}
Y.~Suzumura and R.~Kato, Jpn. J. Appl. Phys. {\bfseries 56}, 05FB02 (2017).

\bibitem{Suzumura_Conduc_2017}
Y.~Suzumura, J. Phys. Soc. Jpn. {\bfseries 86}, 124710 (2017).

\bibitem{Suzumura2018_JPSJ_T}
Y.~Suzumura, H.~Cui, and R.~Kato, J. Phys. Soc. Jpn. {\bfseries 87}, 084702
  (2018).

\bibitem{Suzumura_Yamakage_JPSJ}
Y.~Suzumura and A.~Yamakage, J. Phys. Soc. Jpn. {\bfseries 87}(9), 093704
  (2018).

\bibitem{Kim_Rappe_NLS_2015}
Y.~Kim, B.~J. Wieder, C.~L. Kane, and A.~M. Rappe, Phys. Rev. Lett. {\bfseries
  115}, 036806 (2015).

\bibitem{Weng_Line_node}
H.~Weng, Y.~Liang, Q.~Xu, R.~Yu, Z.~Fang, X.~Dai, and Y.~Kawazoe, Phys. Rev. B
  {\bfseries 92}, 045108 (2015).

\bibitem{KieranPRL}
K.~Mullen, B.~Uchoa, and D.~T. Glatzhofer, Phys. Rev. Lett. {\bfseries 115},
  026403 (2015).

\bibitem{Fang_NLS}
C.~Fang, Y.~Chen, H.-Y. Kee, and L.~Fu, Phys. Rev. B {\bfseries 92}, 081201
  (2015).

\bibitem{Liu_Balents2017}
J.~Liu and L.~Balents, Phys. Rev. B {\bfseries 95}, 075426 (2017).

\bibitem{Katayama2006_JPSJ75}
S.~Katayama, A.~Kobayashi, and Y.~Suzumura, J. Phys. Soc. Jpn. {\bfseries 75},
  054705 (2006).

\bibitem{Herring_PR.52.365}
C.~Herring, Phys. Rev. {\bfseries 52}, 365 (1937).

\bibitem{Fu_Kane_PRB.76.045302}
L.~Fu and C.~L. Kane, Phys. Rev. B {\bfseries 76}, 045302 (2007).

\bibitem{Murakami2007}
S.~Murakami, New J. Phys. {\bfseries 9}, 356 (2007).

\bibitem{Burkov2011}
A.~A. Burkov, M.~D. Hook, and L.~Balents, Phys. Rev. B {\bfseries 84}, 235126 
  (2011).

\bibitem{Hirayama2017}
M.~Hirayama, R.~Okugawa, T.~Miyake, and S.~Murakami,
 Nat. Commun. 8, 14022 (2017).

\bibitem{Hirayama2018}
M.~Hirayama, R.~Okugawa, and S.~Murakami, J. Phys. Soc. Jpn {\bfseries 87},
  041002 (2018).



\bibitem{Yamakage_CaAgX_Dirac}
A.~Yamakage, Y.~Yamakawa, Y.~Tanaka, and Y.~Okamoto, J. Phys. Soc. Jpn.
  {\bfseries 85}, 013708 (2016).

\bibitem{Liu2016PRB}
H.~Huang, J.~Liu, D.~Vanderbilt, and W.~Duan, Phys. Rev. B {\bfseries 93},
  201114 (2016).

\bibitem{Diracloop_CaAs3_2017}
Y.~Quan, Z.~P. Yin, and W.~E. Pickett, Phys. Rev. Lett. {\bfseries 118},
  176402 (2017).

\bibitem{Ag2S_PRB.96.115106}
H.~Huang, K.-H. Jin, and F.~Liu, Phys. Rev. B {\bfseries 96}, 115106 (2017).

\bibitem{Fu_Kane_Mele2007}
L.~Fu, C.~L. Kane, and E.~J. Mele, Phys. Rev. Lett. {\bfseries 98}, 106803
  (2007).

\bibitem{Liu2018}
Z.~Liu, H.~Wang, Z.~F. Wang, J.~Yang, and F.~Liu, Phys. Rev. B {\bfseries 97},
  155138 (2018).

\bibitem{Kino2006_JPSJ}
H.~Kino and T.~Miyazaki, J. Phys. Soc. Jpn. {\bfseries 75}, 034704 (2006).

\bibitem{Koretsune_Hotta_PRB2014}
T.~Koretsune and C.~Hotta, Phys. Rev. B {\bfseries 89}, 045102 (2014).


\bibitem{Luttinger1955}
J. M. Luttinger and W. Kohn, Phys. Rev. {\bf 97}, 869 (1955).

\bibitem{Kobayashi2007_JPSJ}
A.~Kobayashi, S.~Katayama, Y.~Suzumura, and H.~Fukuyama, J. Phys. Soc. Jpn.
  {\bfseries 76}, 034711 (2007).

\bibitem{Wimmer1981}
E.~Wimmer, H.~Krakauer, M.~Weinert, and A.~J. Freeman, Phys. Rev. B {\bfseries
  24}(2), 864 (1981).

\bibitem{L_KA}
D.~D. Koelling and G.~O. Arbman, J. Phys. F, Metal Phys. {\bfseries 5}, 2041
  (1975).

\bibitem{Weinert}
M.~Weinert, J. Math. Phys. {\bfseries 22}, 2433 (1981).

\bibitem{GGA_PBE}
J.~P. Perdew, K.~Burke, and M.~Ernzerhof, Phys. Rev. Lett. {\bfseries 77},
  3865 (1996).

\bibitem{Imp_Tetrahedron1994}
P.~E. Bl\"ochl, O.~Jepsen, and O.~K. Andersen, Phys. Rev. B {\bfseries 49},
  16223 (1994).

\bibitem{Katayama2009_EPL}
S.~Katayama, A.~Kobayashi, and Y.~Suzumura, Eur. Phys. J. B {\bfseries 67},
  139 (2009).


\end{thebibliography}


\end{document}